\documentclass[aps,prl,reprint,superscriptaddress]{revtex4-2}

\usepackage{blindtext}
\usepackage{amssymb}
\usepackage{amsmath}
\usepackage{mathtools}
\usepackage{cases}
\usepackage[normalem]{ulem}
\usepackage{physics}
\usepackage{xcolor}
\usepackage{calc}
\usepackage{hyperref}
\hypersetup{colorlinks, allcolors=blue}
\usepackage[capitalise]{cleveref}
\usepackage{tikz}
\usepackage{comment}

\usetikzlibrary{
    math,
    arrows,
    arrows.meta,
    decorations.pathmorphing, 
    decorations.shapes,
    positioning,
    calc,
    shapes,
    shapes.geometric
}

\newcommand{\subcref}[2]{\cref{#1}\hyperref[#1]{(#2)}}

\renewcommand{\section}[1]{\unskip\textit{#1}---\ignorespaces}

\newcommand{\bexpval}[1]{\big\langle #1 \big\rangle}

\makeatletter
\def\maketitle{
\@author@finish
\title@column\titleblock@produce
\suppressfloats[t]}
\makeatother

\newcommand\titletext{Superbunching from coherently driven atoms in a waveguide}
\newcommand\affc{Department of Microtechnology and Nanoscience (MC2),
Chalmers University of Technology, 41296 Gothenburg, Sweden}
\newcommand\affr{Theoretical Quantum Physics Laboratory, RIKEN Cluster for Pioneering Research, Wako-shi, Saitama 351-0198, Japan}

\begin{document}

\title{\titletext}
\author{Zeidan Zeidan}
\email{zeidan.zeidan@chalmers.se}
\affiliation{\affc}
\author{Therese Karmstrand}
\affiliation{\affc}
\affiliation{\affr}
\author{Maryam Khanahmadi}
\affiliation{\affc}
\author{Göran Johansson}
\affiliation{\affc}
\date{\today}

\begin{abstract}
We investigate the scattered field from $N$ identical two-level atoms resonantly driven by a weak coherent field in a one-dimensional waveguide. For atoms separated by the drive wavelength, increasing the number of atoms progressively suppresses transmission while enhancing photon bunching. Transmission becomes a superbunched $(N+1)$-photon scattering process that is predominantly incoherent. Remarkably, we find that this transmission is only possible through a process where all $N$ atoms are excited, enabling heralded multi-photon state generation with applications in long-distance entanglement and quantum metrology. 
\end{abstract}

\maketitle

\section{Introduction}
Nonlinear light-matter interactions have been studied for decades, leading to the discovery of various quantum phenomena including nonclassical light state generation \cite{Corzo2019, Sheremet2023}, antibunching \cite{Prasad2020, Sheremet2023}, and superradiance \cite{Dicke1954, Gross1982, Goban2015, Sheremet2023}. Multi-photon states, in particular, enable applications in quantum metrology \cite{Paulisch2019, PerarnauLlobet2020, Khanahmadi2023jan}, entanglement generation for quantum communication \cite{Blauensteiner2009, Maffei2024}, and quantum lithography \cite{Boto2000, Bjork2001}.

Waveguide quantum electrodynamics (wQED) has shown to be a promising platform for light-matter interactions due to the strong coupling between atoms and electromagnetic fields propagating in a one-dimensional waveguide. Experimentally, wQED has been demonstrated on various platforms \cite{Sheremet2023, Chang2007}, including single-mode nanofibers coupled to cold atoms \cite{Vetsch2010, Pennetta2022}, quantum dots in photonic crystals \cite{Lodahl2004}, and superconducting circuits \cite{Astafiev2010, Abdumalikov2010, Hoi2011, Hoi2012, Hoi2013, vanLoo2013}.

It is well known that an atom in a waveguide almost perfectly reflects a weak resonant drive, and that the transmitted field is strongly bunched \cite{Ficek2005, Shen2005aug, Shen2005nov, Chang2007, Zheng2010, Astafiev2010, Hoi2011, Hoi2012, Peropadre2013, Hoi2013, Zhang2018}. For a single atom, the bunching occurs, somewhat counterintuitively, because a transmission event predominantly heralds the excitation of the atom \cite{Zhang2018}, which then spontaneously emits a second photon. While this single-atom picture is well established, its extension to two or more atoms in the waveguide remains unexplored.

In this Letter, we study the scattering of a resonant coherent state by $N$ atoms in a waveguide and demonstrate that, when driving weakly, increasing the number of atoms progressively suppresses the transmission while enhancing photon bunching. By considering atoms separated by the resonant wavelength, we derive exact analytical expressions describing the scattering dynamics. We find that transmission in the weak-drive regime is a superbunched $(N+1)$-photon scattering process that is predominantly incoherent, with no phase relation to the drive. Remarkably, we analytically find that this transmission is only possible through a process where all $N$ atoms are collectively excited. Thus, the first detection of a transmitted photon in a bunch heralds the atomic system most likely in a fully excited state. We verify our analytical findings with numerical simulations. Naturally, this simple setup serves as an entanglement source and enables heralded multi-photon state generation with applications in quantum metrology.

\section{Model}
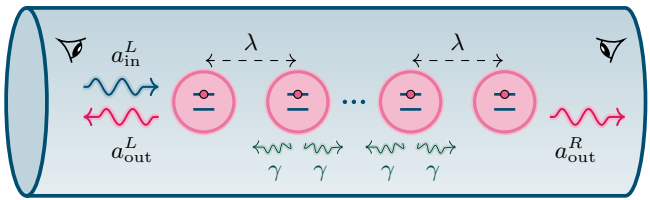
\begin{figure}
    \centering
    \newcommand*\lateraleyee{
    \begin{tikzpicture}[x=0.75pt, y=0.75pt, yscale=-0.15, xscale=0.15]
    \draw[very thin]  (300,100.33) .. controls (326,122) and (352,135) .. (378,139.33) .. controls (352,143.67) and (326,156.67) .. (300,178.33) ;
    \fill (308.94,116.33) .. controls (313.87,116.33) and (317.86,125.51) .. (317.85,136.83) .. controls (317.84,148.15) and (313.84,157.33) .. (308.91,157.33) .. controls (303.99,157.32) and (300,148.14) .. (300.01,136.82) .. controls (300.02,125.5) and (304.02,116.32) .. (308.94,116.33) -- cycle ;
    \draw[very thin] (314.84,166.6) .. controls (311.87,164.64) and (309.14,162.18) .. (306.76,159.24) .. controls (295.12,144.82) and (296.6,124.33) .. (310.07,113.45) .. controls (311.48,112.32) and (312.96,111.33) .. (314.5,110.49) ; 
    \fill[white] (304.43,124.2) .. controls (306.09,124.25) and (307.32,128.01) .. (307.18,132.6) .. controls (307.05,137.19) and (305.59,140.88) .. (303.93,140.83) .. controls (302.27,140.78) and (301.03,137.02) .. (301.17,132.43) .. controls (301.31,127.83) and (302.76,124.15) .. (304.43,124.2) -- cycle ;
    \end{tikzpicture}
}

\newcommand*\lateraleye{
\begin{tikzpicture}
    %Eyelid
    \draw[black, line width=25pt]  
        (0, 15) .. controls (4, 10) and (8, 7.5) .. (12, 7) 
        .. controls (8, 6.5) and (4, 4) .. (0, 0);
    \draw[black, line width=25pt] 
        (1.8, 1.85) .. controls (2, 1.5) and (1.5, 2) .. (0.9, 2.8) 
        .. controls (-1.5, 6) and (-1.3, 10) .. (1.7, 12) 
        .. controls (2,12. 2) and (2.3, 12.4) .. (2.3, 12.4);
    %Iris
    \fill[black] (1.2, 7.3) ellipse (1.5 and 4);
    \fill[white] (0.2, 8.3) ellipse (0.55 and 1.6);
\end{tikzpicture}
}

\begin{tikzpicture}

\tikzset{
    glow/.style={
        preaction={
            draw,
            #1!30,
            line width=2.5pt,
            -
        },
        preaction={
            draw,
            #1!10,
            thick
        }
    }
}

\tikzset{
    glowsmall/.style={
        preaction={
            draw,
            #1!30,
            line width=1.5pt,
            -
        },
        preaction={
            draw,
            #1!10,
            thick
        }
    }
}

\tikzset{cylinder end fill/.style={path picture={
\pgftransformshift{\centerpoint}
\pgftransformrotate{\rotate}
\pgfpathmoveto{\beforetop}
\pgfpatharc{90}{-270}{\xradius and \yradius}
\pgfpathclose
\pgfsetfillcolor{#1}
\pgfusepath{fill}}
}}

\definecolor{wg}{RGB}{0,78,114}
\definecolor{other}{RGB}{0,77,64}
\definecolor{pink}{RGB}{255,228,223}
\definecolor{atom}{RGB}{216,27,96}

% Waveguide
\node[
cylinder, 
line width=1.4pt,
cylinder end fill=wg!25,
left color=wg!25,
middle color=wg!15,
right color=wg!10,
shading angle=0,
%cylinder body fill=wg!10,
shape border rotate=180, 
color=wg,
draw,
minimum height=8.5cm,
minimum width=2.5cm,
shape aspect=2.5,
] (A) {};

% Atoms
\tikzmath{
    \centery=1.5;
    \a1=2.5-0.125;
    \a2=3.5+0.125;
    \a3=5+0.25-0.125;
    \a4=6+0.25+0.125;
}

\begin{scope}[xshift=-4.2cm, yshift=-\centery cm]
\foreach \x in {\a1,\a2,\a3,\a4} {
    \filldraw[glow=atom, color=atom!60, fill=atom!30, very thick](\x,\centery) circle (0.4); % Atom outer circle
    \draw[color=wg, line width=1pt] (\x-0.14,\centery cm+0.1cm) -- (\x+0.14,\centery cm+0.1cm); %TLS line top
    \draw[color=wg, line width=1pt] (\x-0.14,\centery cm-0.1cm) -- (\x+0.14,\centery cm-0.1cm); %TLS line bottom
    \node[draw, circle,fill=atom!70,scale=0.35] at (\x,\centery cm+0.1cm){};
}

\draw [line width=0.6mm, line cap=round, color=wg, dash pattern=on 0pt off 2\pgflinewidth] (\a2 cm+0.5cm+0.125cm,\centery cm) -- (\a3 cm-0.5cm-0.125cm,\centery cm);

\foreach \x in {\a1,\a3} {
    \draw (\x cm+0.5cm+0.125cm,\centery cm + 0.8cm) node[] {$\lambda$};
    \draw[<->, dashed, black] (\x+0.025,\centery cm+0.55cm) to (\x+1.25-0.025,\centery cm+0.55cm);
}

\draw[->, glow=wg, thick, wg, decorate, decoration={snake,amplitude=3pt,pre length=1.5pt,post length=2.5pt}] (\a1 cm - 1.6cm,\centery+0.2) -- node[above=0.15cm, xshift=0.15cm]{$\textcolor{black}{a_\text{\makebox[\widthof{out}][l]{in}}^L}$} ++(1,0);
\draw[<-, thick, atom, decorate, decoration={snake,amplitude=3pt,pre length=5.5pt,post length=0pt}, glow=atom] (\a1 cm - 1.6cm,\centery-0.2) -- node[below=0.15cm, xshift=0.15cm]{$\textcolor{black}{a_\text{out}^L}$} ++(1,0);

\draw[->, glow=atom, thick, atom, decorate, decoration={snake,amplitude=3pt,pre length=1.5pt,post length=2.5pt}] (\a4 cm + 0.6cm,\centery-0.2) -- node[below=0.15cm, xshift=-0.15cm]{$\textcolor{black}{a_\text{out}^R}$} ++(1,0);

\foreach \x in {\a2,\a3} {
\draw (\x+0.3,\centery-0.95) node[] {$\textcolor{other}{\gamma}$};
\draw (\x-0.3,\centery-0.95) node[] {$\textcolor{other}{\gamma}$};
\draw[->, glowsmall=other, other, decorate, decoration={snake,segment length=4pt,amplitude=1.5pt,post length=5pt}] (\x-0.1,\centery-0.55) to [out=225, in=160](\x-0.6,\centery-0.6);
\draw[->, glowsmall=other, other, decorate, decoration={snake,segment length=4pt,amplitude=1.5pt,post length=5pt,mirror}] (\x+0.1,\centery-0.55) to [out=-45, in=15](\x+0.6,\centery-0.6);
}

% Detector
\tikzmath{
    \detectorr=0.18;
}
\node[scale=0.025, rotate=30] at (7.8, 2.25) {\lateraleye};
\node[scale=0.025, rotate=-30, xscale=-1] at (0.6, 2.25) {\lateraleye};

\end{scope}

\end{tikzpicture}
\caption{\label{fig:waveguide}$N$ identical atoms, modeled as two-level systems, coupled to a waveguide. The atoms are spaced apart by wavelength distance $\lambda$ and decay to the left and right propagating fields with spontaneous emission rate $\gamma$. The operator $a_\text{in}^L$ refers to the annihilation operator of the incoming coherent drive from the left and $a_\text{out}^{R/L}$ is the operator of the outgoing field propagating toward the right/left. The eyes in the two corners represent photon detectors.
}
\end{figure}

We consider $N$ identical atoms spaced one radiation wavelength apart in a one-dimensional waveguide. The atoms are modeled as two-level systems with resonance frequency $\omega$ and are symmetrically coupled to the waveguide into which they decay through spontaneous emission with the rate $\gamma$ to the left and right propagating fields (see \cref{fig:waveguide}). 
Neglecting the time delay between the closely spaced atoms, the master equation for the density matrix $\rho$ of the atomic system, collectively and coherently driven with Rabi frequency $\Omega$, is
\begin{equation}
    \label{eq:master}
    \dot{\rho} = -i \tfrac{\Omega}{2}[S_+ + S_-, \rho] + \mathcal{D}(a_\text{out}^R)\rho + \mathcal{D}(a_\text{out}^L) \rho
\end{equation}
where $\hbar=1$, $\mathcal{D}(L)\rho = L\rho L^\dag - \frac{1}{2}\{L^\dag L, \rho\}$ is the Lindblad dissipator, $S_\pm = \sum_{i=1}^N S_{\pm}^i$ are the collective atomic dipole operators, and $a_\text{out}^{R/L}$ is the annihilation operator of the outgoing field propagating toward the right/left. The outgoing field operators are related to the annihilation operators of the incoming field from the right/left, $a_\text{in}^{R/L}$, through the input-output relations $a_\text{out}^L = a_\text{in}^R + \sqrt{\gamma}S_-$ and $a_\text{out}^R = a_\text{in}^L + \sqrt{\gamma}S_-$ \cite{Lalumire2013, supp}\nocite{Riordan1968, Mitzenmacher2005}. We consider an incoming coherent drive from the left with rate amplitude $\alpha = i\Omega/\sqrt{\gamma}$, and vacuum incoming from the right. When individual atomic relaxation and time delays are negligible, an atomic system initialized in the ground state evolves within the symmetric $(N+1)$-dimensional Dicke subspace \cite{Dicke1954}. In the Hilbert space of the atomic system, the input-output relations then become $a_\text{out}^L = \sqrt{\gamma}S_-$ and $a_\text{out}^R = \alpha I_{N+1} + \sqrt{\gamma}S_-$, where $I_{N+1}$ is the identity operator. Using the input-output relations, we can rewrite \cref{eq:master} as $\dot{\rho} = 2\mathcal{D}(a_\text{out}^R)\rho$, to immediately obtain the steady-state solution \cite{Puri1979, Puri1980, Drummond1980, Lawande1981, Hassan1980, Hassan1982}:
\begin{equation}
\label{eq:steady_state}
    \rho = \frac{(a_\text{out}^{R\dag}a_\text{out}^R)^{-1}}{\Tr \big(a_\text{out}^{R\dag}a_\text{out}^R\big)^{-1}} = \frac{1}{D} \sum_{m,n=0}^N  \left(\frac{S_-}{g^*}\right)^m\left(\frac{S_+}{g}\right)^n,
\end{equation}
where the normalized drive amplitude $g=\alpha/\sqrt{\gamma}=i\Omega/\gamma$ and the normalization constant $D=\sum\limits_{r=0}^{N} \binom{N+r+1}{2r+1} (r!)^2 |g|^{-2r}$. Using the steady-state solution from \cref{eq:steady_state}, we now calculate the scattered photon rates, providing initial insight into the system's dynamics and photon bunching behavior.

\section{Scattered field}
The average photon rate in the reflected and transmitted directions is given by $\bar{n}_\text{ref} = \bexpval{a_\text{out}^{L\dag} a_\text{out}^L}$ and $\bar{n}_\text{trans} = \bexpval{a_\text{out}^{R\dag} a_\text{out}^R} $, respectively. The scattered photon rate comprises phase-coherent components $\bar{n}_\text{ref,coh} = \left|\bexpval{a_\text{out}^{L}}\right|^2 $ and $\bar{n}_\text{trans,coh} = \left|\bexpval{a_\text{out}^{R}}\right|^2$, with the remaining phase-incoherent components given by $\bar{n}_\text{ref/trans,inc} = \bar{n}_\text{ref/trans}-\bar{n}_\text{ref/trans,coh}$. The scattered phase-incoherent photons, attributed to spontaneous emission, are equal in the reflected and transmitted directions, $\bar{n}_\text{ref,inc} = \bar{n}_\text{trans,inc}$, due to the symmetry of emission into the waveguide. When weakly driven, $|g| \ll 1$, the atoms collectively oscillate coherently with a phase opposite to the input drive, leading to near-perfect reflection and effectively canceling the transmitted field \cite{supp}, as in the single-atom case \cite{Ficek2005, Shen2005aug, Shen2005nov, Chang2007, Zheng2010, Astafiev2010, Hoi2011, Hoi2012, Peropadre2013, Hoi2013, Zhang2018}. Notably, in this regime, the transmission is predominantly incoherent and is given by \cite{supp}:
\begin{equation}
\label{eq:weak_transmitted_power}
    \bar{n}_\text{trans} \approx \bar{n}_\text{trans,inc} = \gamma |g|^{2(N+1)} \frac{N+1}{(N!)^2} + \mathcal{O}\big(|g|^{2(N+2)}\big),
\end{equation}
which is proportional to the incoming rate of $(N+1)$-photon states from the Poissonian drive. The scaling in \cref{eq:weak_transmitted_power} reveals that transmission requires driving the atomic system with sufficient photons to excite all $N$ atoms. This behavior can be understood through two perspectives: (1) $N$ atoms can coherently reflect Fock states of up to $N$ photons, while states with $N+1$ or more photons enable incoherent transmission. (2) Although the average drive amplitude is in the linear regime, amplitude fluctuations occasionally probe the atomic nonlinearity. 

Moreover, in the weak-drive regime, the probability density for simultaneous detection of $n$ transmitted photons is proportional to the zero-time $n$-th order correlation function \cite{supp}:
\begin{equation}
\label{eq:weak_transmitted_Gn}
    G^{(n)}_\text{trans}(0) = 
    \begin{cases*}
        \gamma^n |g|^{2(N+1)} \tbinom{N+n}{2n-1} \left[\tfrac{(n-1)!}{N!}\right]^2 & if $n \leq N$, \\
        \gamma^{n} |g|^{2n} \tbinom{n-1}{N}^2 & if $n > N$.
    \end{cases*}
\end{equation}
For $n \leq N+1$, $G^{(n)}_\text{trans}(0)$ scales as $|g|^{2(N+1)}$, matching the probability density of simultaneously detecting $N+1$ photons from the drive. In contrast, for $n>N+1$, the $n$-th order correlation function of both the transmitted field and input drive scale as $|g|^{2n}$. This scaling highlights that transmission is a process unlocked by at least $N+1$ input photons from the drive. While transmission events are rare in the weak driving regime, the second-order coherence function $g^{(2)}_\text{trans}(0) \propto |g|^{-2(N+1)} \gg 1$ reveals superbunched \cite{Ficek2005} transmission when it occurs. To further understand this $(N+1)$-photon transmission process, we next examine the atomic system's quantum state conditioned on transmission.

\section{Conditional atomic state}
When conditioning the steady state from \cref{eq:steady_state} on detecting a transmitted photon, we obtain the maximally mixed state
\begin{equation}
    \label{eq:ss_conditional}
    \rho_\text{c}=\frac{a^R_\text{out} \rho a_\text{out}^{R\dag}}{\bexpval{ a_\text{out}^{R\dag} a^R_\text{out}}}=\frac{I_{N+1}}{N+1}.
\end{equation}
In the strong driving regime, $|g| \gg 1$, the conditional state in \cref{eq:ss_conditional}, to leading order, coincides with the steady state in \cref{eq:steady_state}, as the drive saturates the atomic system and photon detections have a negligible effect on the state. However, the conditional state in \cref{eq:ss_conditional} is \emph{independent} of the normalized drive amplitude $g$. This raises the question of how to understand this state in the weak driving regime, $|g| \ll 1$, where the steady state is close to the ground state. In this regime, while the probability of detecting a transmitted photon is small, proportional to $|g|^{2(N+1)}$, when such a photon is detected, \cref{eq:ss_conditional} tells us that the atomic system has an average of $N/2$ excitations. This can be understood by highlighting the fact that we are conditioning on detecting \emph{any} transmitted photon. Transmission only occurs in bunches, as seen from \cref{eq:weak_transmitted_power,eq:weak_transmitted_Gn}, following the collective excitation of all atoms, and each photon in the bunch could have been emitted from any excitation level of the system. This lack of information about which excitation level the transmitted photon was emitted from leads to maximal uncertainty in the atomic state.

To determine the atomic state after a specific transmitted click, we turn to the normalized no-click master equation, a master equation with removed jump/click terms \cite{supp, Molmer1993, Breuer2007, Wiseman2009}. In the weak driving regime, the photon scattering rate is small, justifying the study of the no-click master equation. First, we consider the evolution described in \cref{eq:master} without transmitted clicks by removing the term $a^R_\text{out} \rho a_\text{out}^{R\dag}$. When we condition the steady state of this no-click evolution on an inevitable transmitted detection, we obtain the atomic state after the \emph{first} transmitted click in a photon bunch. Using $\ket{k}_\text{ex}$ to denote the state with $k$ excited atoms, this first-click state is given by \cite{supp}:
\begin{equation}
\label{eq:ss_conditional_no_transmitted_clicks}
    \rho_\text{c}^\text{first} = \sum_{k=0}^N \frac{1}{2^{N-k}(2-2^{-N})} 
    \ketbra{k}_\text{ex}.
\end{equation}
Notably, the state in \cref{eq:ss_conditional_no_transmitted_clicks} follows an $N$-truncated geometric distribution, arising from the possibility of $N-k$ reflected photons preceding the first transmitted photon. This distribution is consistent with all incoherent scattering events involving $N+1$ photons, each having an equal $1/2$ probability of transmission or reflection. The highest probability corresponds to the fully excited state, i.e., $k=N$, occurring when the first photon is transmitted, and decreases by half with each additional previously reflected photon. The geometric distribution reveals that transmission originates from the fully excited state. We thus consider \cref{eq:master} without both reflected and transmitted clicks by removing the jump terms $a^L_\text{out} \rho a_\text{out}^{L\dag}$ and $a^R_\text{out} \rho a_\text{out}^{R\dag}$. Indeed, conditioning the steady state of this no-click evolution on an inevitable transmitted detection yields the fully excited atomic state $\ket{N}_\text{ex}$, confirming that transmission occurs through a process where all $N$ atoms are excited.

\section{Timescales}
Having established that transmission is a multi-photon process, enabled when $(N+1)$ or more photons scatter off the atoms, we now make this picture more quantitative by defining the timescale $T_\text{in}$ within which these photons must arrive to enable this transmission. In the weak-drive regime, transmission is dominated by the incoming rate of $N+1$ photons, as shown in \cref{eq:weak_transmitted_power}. Incoming photons follow a Poisson distribution, and under weak driving, the probability of $N+1$ photons arriving within time $T_\text{in}$ is $(\gamma|g|^2 T_\text{in})^{N+1}/(N+1)!$. These incoming photons yield $(N+1)/2$ transmitted photons, as they can either be reflected or transmitted, resulting in an average transmission rate of $(\gamma|g|^2)^{N+1}T_\text{in}^{N}/(2N!)$ \cite{supp}. By equating this rate with the transmission rate from \cref{eq:weak_transmitted_power}, we arrive at
\begin{equation}
\label{eq:T_in}
    T_\text{in} = \frac{1}{\gamma} \left(\frac{2(N+1)}{N!}\right)^{1/N}.  
\end{equation}
\cref{eq:T_in} shows that, with increasing $N$, more input photons are required within a shorter time frame to trigger transmission. For large $N$, $T_\text{in}$ is lower-bounded by $e/(\gamma N)$, qualitatively agreeing with the increasing collective coupling of the atomic system. 

The second relevant timescale, $T_\text{out}$, characterizes the relaxation of an excited atomic state. From the fully excited state $\ket{N}_\text{ex}$, the atomic system undergoes cascaded emission from $\ket{k}_\text{ex}$ to $\ket{k-1}_\text{ex}$ at rates $\Gamma_{k} = 2\gamma \mel{k}{S_+ S_-}{k}_\text{ex} = 2\gamma(N-k+1)k$, until reaching the ground state $\ket{0}_\text{ex}$. This process forms a continuous-time absorbing Markov chain with $N+1$ states, where decay times between adjacent states follow exponential distributions with rates $\Gamma_k$. The total relaxation time follows a hypoexponential distribution \cite{Amari1997, Bolch2006, Legros2015}, which is the sum of independent exponential random variables, with rate parameters $\Gamma_N, \dots, \Gamma_1$. Crucially, this statistical framework enables calculation of the $p$-th relaxation time quantile, $T_{\text{out},p}$, defined as the time by which the atomic system reaches the ground state with probability $p$. Additionally, the average relaxation time \cite{Gross1982, supp} is given by $T_\text{out,avg} = \sum_{k=1}^{N} \frac{1}{\Gamma_k} = \frac{H_N}{\gamma(N+1)}$, where $H_N$ is the $N$-th Harmonic number. The asymptotic limit for large $N$ is $\ln(N)/(\gamma N)$, characteristic of superradiant collective spontaneous emission \cite{Gross1982}. With these timescales established, we now turn to photon detection statistics obtained by numerical simulations.
\begin{figure}
    \centering
    \includegraphics{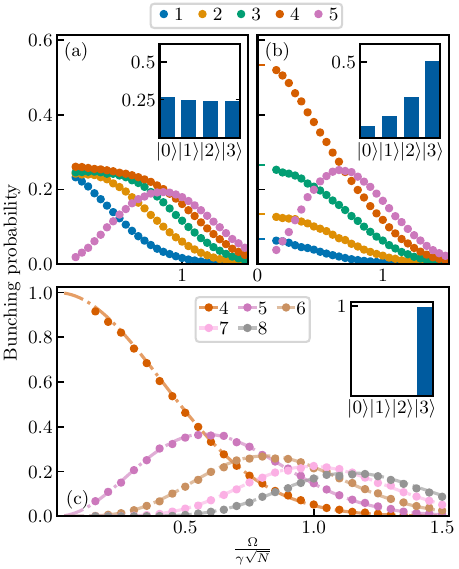}
    \caption{Bunching probability versus effective drive amplitude $\Omega/(\gamma \sqrt{N})$, with $N=3$ atoms, $\gamma=1$, $T_\text{in} = 1.10$ according to \cref{eq:T_in}, and $T_\text{out} = T_\text{out,99\%} = 1.30$. The circles show simulation results, and different colors represent different photon bunch sizes. The insets display the average atomic state populations, conditioned on the first transmitted click, for the simulations with the weakest drive strength. For stronger drive strengths, the probabilities sum to less than one as the curves for larger-sized bunches are omitted. (a) All transmitted photons are counted as first clicks, i.e., clicks are recounted. (b) The first transmitted click is defined as a transmitted detection preceded by no other transmitted photons during the time $T_\text{in}$. (c) The first transmitted click is defined as a transmitted detection preceded by neither transmitted nor reflected photons during the time $T_\text{in}$. The dash-dotted lines represent analytical predictions from \cref{eq:photon_bunch_prob} with the fitted timescales $\hat{T}_\text{in}=0.37$ and $\hat{T}_\text{out}=0.93$.}
    \label{fig:photon_bunches}
\end{figure}

\section{Photon detection statistics}
The master equation in \cref{eq:master} is simulated using Monte Carlo trajectory evolution \cite{Molmer1993, Lambert2024}, with the output operators $a_\text{out}^{R/L}$ corresponding to photon detection in the transmitted and reflected fields according to \cref{fig:waveguide}. Here we present results for $N=3$ atoms, while results for $N=1\text{ and }2$ atoms can be found in the End Matter. For our analysis, a photon bunch is defined by the number of clicks in the reflected and transmitted fields within a bunching time $T_\text{out}$ after, and including, the first transmitted click. \cref{fig:photon_bunches} shows the probability of obtaining a photon bunch of a given size versus the effective drive amplitude $\Omega/(\gamma \sqrt{N})$, demonstrating that different definitions of the \emph{first transmitted click} yield different photon bunching statistics. 

In \subcref{fig:photon_bunches}{a}, we employ a simple counting method where every transmitted photon is treated as a first click, effectively recounting photons that belong to the same bunching event. In the weak-drive regime we detect the mixed state as predicted in \cref{eq:ss_conditional}, however, for larger drive strengths most of the detected photons originate directly from the coherent drive and bunches typically contain more than five photons. In \subcref{fig:photon_bunches}{b}, we instead define the first transmitted click as a transmitted detection preceded by no other transmitted photons within the characteristic timescale $T_\text{in}$. In the weak-drive regime, the bunching probabilities approach the truncated geometric distribution, as predicted in \cref{eq:ss_conditional_no_transmitted_clicks}.

In \subcref{fig:photon_bunches}{c}, we define the first transmitted click as a transmitted detection preceded by neither transmitted nor reflected photons during the time $T_\text{in}$. In the weak-drive regime, we approach exclusively $(N+1)$-photon bunches, as predicted by the no-click master equation. The analytical prediction for a $k$-photon bunch probability \cite{supp}, plotted in \subcref{fig:photon_bunches}{c}, is given by
\begin{equation}
\label{eq:photon_bunch_prob}
    p(k) = \sum_{i=0}^{k-N-1} \Pr(Y=k-i)\Pr(Z=i),
\end{equation}
where $k \geq N+1$. The random variable $Y$, representing the initial photons that trigger transmission, follows an $N$-truncated Poisson distribution, while $Z$, accounting for additional input photons arriving during the emission process, follows a Poisson distribution. The summation in \cref{eq:photon_bunch_prob} captures all possible combinations of initial and additional photons that result in a $k$-photon bunch. For example, an $(N+2)$-photon bunch arises from either $N+2$ initial photons with no additional arrivals, or $N+1$ initial photons plus one additional arrival. The random variables $Y$ and $Z$ have the parameters $|\alpha|^2 \hat{T}_\text{in}$ and $|\alpha|^2 \hat{T}_\text{out}$, respectively, where the timescales $\hat{T}_\text{in}$ and $\hat{T}_\text{out}$ are obtained by fitting \cref{eq:photon_bunch_prob} to the simulation data. Interestingly, while these fitted timescales are systematically found to be shorter than those used to define photon bunches in the simulations \cite{supp}, the analytical prediction of \cref{eq:photon_bunch_prob}, nonetheless, shows excellent agreement with the numerical data in \subcref{fig:photon_bunches}{c}, even at intermediate and strong drive. In the weak driving regime, the probability of obtaining an $(N+1)$-photon bunch from \cref{eq:photon_bunch_prob} can be approximated by $p(N+1) = e^{-\Omega^2T/\gamma}$, showing an exponential convergence to $(N+1)$-photon bunches \cite{supp}.
\begin{figure}
    \centering
    \includegraphics{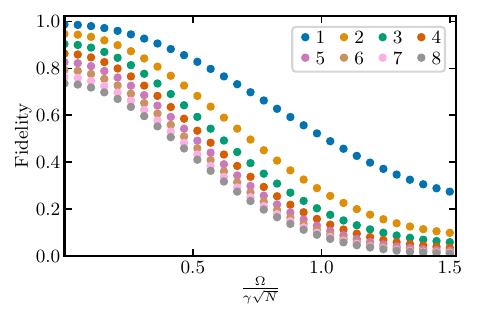}
    \caption{Fidelity of the output field with the entangled binomial state, versus effective drive amplitude $\Omega/(\gamma \sqrt{N})$, for a system of initially fully excited atoms. for $N=1$ to $8$ atoms, represented by different colored markers, and $\gamma=1$. The output states are taken at time $T_\text{out,99\%}$.}\label{fig:binom_output_fidelity}
\end{figure}

\section{Applications}
Naturally, our simple setup enables multi-photon state generation and entanglement distribution, heralded by the detection of the first transmitted photon. Specifically, the bidirectional spontaneous emission from fully excited atoms produces an entangled binomial state, $\ket{\psi_\text{b}}=\sum\limits_{k=0}^N\left[\binom{N}{k}2^{-N}\right]^{1/2}\ket{k,N-k}_\text{L,R}$, between the left and right propagating modes. \cref{fig:binom_output_fidelity} shows the fidelity $\bra{\psi_\text{b}}\varrho_{\text{out}}\ket{\psi_\text{b}}$ 
between the output state $\varrho_\text{out}$ and the binomial state versus the effective drive amplitude $\Omega/(\gamma\sqrt{N})$ for up to eight atoms. Following \cite{Kiilerich2019, Khanahmadi2023oct}, we catch the output state by filtering with the optimal output mode shape corresponding to fully excited atoms decaying without drive. As shown in \cref{fig:binom_output_fidelity}, the fidelity decreases with increased drive amplitude, as atomic emission is superposed with the coherent drive. Furthermore, lower fidelities correspond to higher atom numbers, as nonlinear decay generates multimode output states \cite{Khanahmadi2023oct}. However, the optimal mode nevertheless captures at least 90$\%$ of the output photons for up to eight atoms \cite{PerarnauLlobet2020}. Moreover, placing a mirror at an appropriate distance from the atoms in the transmitted direction \cite{Sathyamoorthy2016} channels all $N$ photons to propagate in a single direction. This enables applications in quantum metrology for quantum-enhanced phase measurements \cite{Paulisch2019, PerarnauLlobet2020, Khanahmadi2023jan} and quantum lithography to beat the diffraction limit \cite{Boto2000, Bjork2001}. 

\section{Discussion}
In summary, we have demonstrated that coherently driven atoms in a waveguide generate superbunched light, with the first transmitted photon heralding the atoms in a fully excited state. In a wider scope, by applying the no-click master equation, we reveal the multi-photon emission mechanisms enabling heralded multi-photon state generation. While we study an arbitrary number of atoms, we note that the effects of the Markov approximation on photon bunching warrant investigation in future work. Finally, as we envision experimental implementation on superconducting waveguides, we plan to extend this work to transmon systems while studying how anharmonicity, dephasing, and detuning affect the photon bunching robustness.

\section{Acknowledgements}
We thank I. Lyngfelt and B. Bertin-Johannet for feedback on the manuscript. We acknowledge support from the Knut and Alice Wallenberg Foundation through the Wallenberg Centre for Quantum Technology (WACQT). Z. Z. and G. J. acknowledge support from the Swedish Research Council, grant no. 2021-04037. T. K. acknowledges support from the FY2024 JSPS Postdoctoral Fellowship for Research in Japan. The simulations were enabled by resources provided by the National Academic Infrastructure for Supercomputing in Sweden (NAISS), partially funded by the Swedish Research Council through grant no. 2022-06725.

\bibliographystyle{apsrev4-2}
%\bibliography{references.bib}

\onecolumngrid
\vspace{1em}
\begin{center}
{\large \textbf{End Matter}}
\end{center}
\vspace{1em}

\begin{figure*}[htbp]
    \centering
    \includegraphics{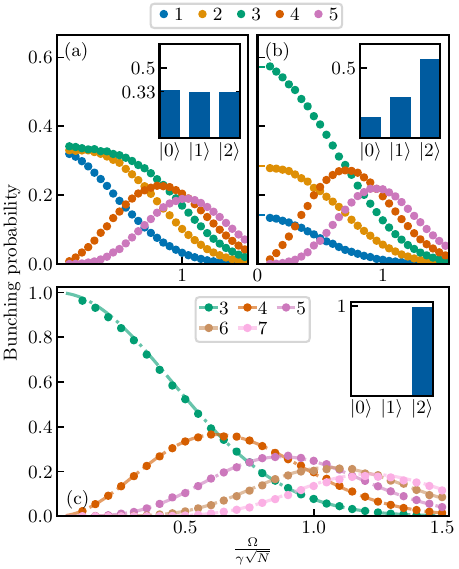}
    \hspace{40pt}
    \includegraphics{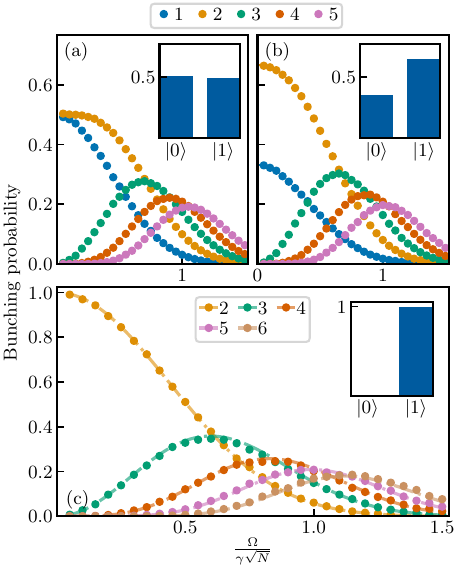}
    \caption{Bunching probability versus effective drive amplitude $\Omega/(\gamma \sqrt{N})$. The left (right) panel corresponds to $N=2\;(1)$ atoms, with the simulation parameters $\gamma=1$, $T_\text{in} = 1.70$, $T_\text{out} = T_\text{out,99\%} = 1.66$, and $\gamma=1$, $T_\text{in} = 4$, $T_\text{out} = T_\text{out,99.9\%} = 3.45$, respectively. The circles show simulation results, and different colors represent different photon bunch sizes. The insets display the average atomic state populations, conditioned on the first transmitted click, for the simulations with the weakest drive strength. For stronger drive strengths, the probabilities sum to less than one as the curves for larger-sized bunches are omitted. (a) All transmitted photons are counted as first clicks, i.e., clicks are recounted. (b) The first transmitted click is defined as a transmitted detection preceded by no other transmitted photons during the time $T_\text{in}$. (c) The first transmitted click is defined as a transmitted detection preceded by neither transmitted nor reflected photons during the time $T_\text{in}$. The dash-dotted lines represent analytical predictions from \cref{eq:photon_bunch_prob} with the fitted timescales $\hat{T}_\text{in}=0.83$, and $\hat{T}_\text{out}=1.05$ for the left panel, and $\hat{T}_\text{in}=2.58$, and $\hat{T}_\text{out}=1.80$ for the right panel.}
    \label{fig:combined}
\end{figure*}
\end{document}

\title{\large Supplemental Material:\texorpdfstring{\\}{} \titletext}

\author{Zeidan Zeidan}
\affiliation{\affc}
\author{Therese Karmstrand}
\affiliation{\affc}
\affiliation{\affr}
\author{Maryam Khanahmadi}
\affiliation{\affc}
\author{Göran Johansson}
\affiliation{\affc}
\date{\today}

\maketitle
\onecolumngrid

This supplemental material provides detailed derivations of the main text results. \cref{ssec:intro,ssec:intro_waveguide} give background for Eqs. (1) and (2). Derivations for the results presented in the main text are located as follows: the transmission rate of Eq. (3) in \cref{ssec:scattered_field}; the higher order correlation function of Eq. (4) in \cref{ssec:scattered_field_higher_trans_weak}; the conditional state of Eq. (6) in \cref{ssec:no_click_trans}; the timescale of Eq. (7) in \cref{ssec:timescale_in}; and the photon bunching probability of Eq. (8) in \cref{ssec:photon_bunches_dist}.

\section{Collectively driven atomic system}
\label{ssec:intro}
The Lindblad master equation for an on-resonance collectively driven atomic system in a frame rotating with the drive frequency, also known as the driven Dicke model \cite{Puri1979, Puri1980, Drummond1980, Hassan1980, Lawande1981, Hassan1982}, is obtained as ($\hbar=1$)
\begin{equation}
\label{seq:master}
    \dot{\rho} = -i\Omega [S_+ + S_-, \rho] + 2\gamma (S_- \rho S_+ - \tfrac{1}{2}\{S_+S_-, \rho\}),
\end{equation}
where $2\gamma$ and $\Omega$ correspond to the spontaneous decay rate for a single atom and the Rabi frequency of the coherent drive, respectively. The collective atomic dipole operators $S_\pm = \sum_{i=1}^N S_{\pm,i}$, for $N$ atoms, satisfy the following angular momentum commutation relations:
\begin{subequations}
\begin{align}
        [S_+,S_-] &= 2S_z, \\
        [S_z,S_\pm] &= \pm S_\pm,
\end{align}
\end{subequations}
with $S_z$ being the collective atomic energy operator. Due to the system's collective behavior, the operators are determined in an $(N+1)$-dimensional Hilbert space. For the derivations in this text, we follow Puri and Lawande \cite{Puri1979} and use the collective atomic states $\ket{p}$, where $p \in \{ 0, 1, \dots, N \}$ is the number of atoms in the ground state. While, for clarity in the Letter, we use the atomic states $\ket{N-p}_\text{ex}$ representing the number of atoms in the excited state. The collective atomic dipole operators act on the atomic states as follows:
\begin{subequations}
\label{seq:dipole_operators}
\begin{align}
    S_+ \ket{p} &= \sqrt{p(N-p+1)}\ket{p-1}, \\
    S_- \ket{p} &= \sqrt{(p+1)(N-p)}\ket{p+1}.
\end{align}
\end{subequations}
Although the operator $S_+$ lowers the atomic state quantum number, it acts as a raising operator since the energy of the atomic system is increased. The steady state of the master equation in \cref{seq:master} is \cite{Puri1979, Puri1980, Drummond1980, Hassan1980, Lawande1981, Hassan1982}:
\begin{equation}
\label{seq:steady_state}
\rho = \frac{1}{D} \sum_{m,n=0}^N  \left(\frac{S_-}{g^*}\right)^m\left(\frac{S_+}{g}\right)^n,
\end{equation}
with the normalized drive amplitude $g=i\Omega/\gamma$ and normalization constant $D=\sum_{r=0}^N H_{r} |g|^{-2r}$ where $H_{r} = \frac{(N+r+1)!(r!)^2}{(N-r)!(2r+1)!} = \binom{N+r+1}{2r + 1} (r!)^2$. According to the steady-state solution, we can calculate the correlation functions of the collective atomic dipole operators \cite{Puri1979, Puri1980, Drummond1980, Hassan1980, Lawande1981, Hassan1982}, which we use in the following sections:
\begin{subequations}
\label{seq:avg_observables}
\begin{align}
    \bexpval{S_+} &= \bexpval{S_-}^* = g\left(1-\frac{N+1}{D}\right), \label{seq:avg_S+} \\
    \bexpval{S_+ S_-} &= |g|^2 \left(1-\frac{N+1}{D}\right), \label{seq:avg_S+S-}\\
    \bexpval{S_+^n S_-^n} &= \frac{|g|^{2n}}{D} \sum_{r=n}^N |g|^{-2r} H_{r}, \quad n \leq N, \label{seq:avg_S+nS-n}\\
    \bexpval{S_+^n S_-^m} &= \frac{1}{D} \sum_{r=\text{max}(n,m)}^N (g)^{-r+n} (g^*)^{-r+m} H_{r}, \quad n,m \leq N. \label{seq:avg_S+nS-m}
\end{align}
\end{subequations}
The normalization constant $D$ can be approximated in two different regimes of the dimensionless parameter $|g|$, corresponding to weak and strong external field driving, respectively:
\begin{subnumcases}{}
|g| \ll 1 \quad \Longrightarrow \quad D \sim \frac{(2N+1)!(N!)^2}{(2N+1)!}|g|^{-2N} = (N!)^2 |g|^{-2N}, \label{seq:D_weak_drive}\\
|g| \gg 1 \quad \Longrightarrow \quad D \sim \frac{(N+1)!}{N!} = (N+1). \label{seq:D_strong_drive}
\end{subnumcases}
It is worth noting that, the expectation values in \cref{seq:avg_observables}, and all other expectation values, are calculated using the steady state of the atomic system from \cref{seq:steady_state} unless otherwise stated.

\section{Atomic system in a waveguide}
\label{ssec:intro_waveguide}

As mentioned in the main text, the total system consists of atoms coupled to a waveguide. The atoms are separated by wavelength distances and spontaneously emit to the left or right direction. Assuming a coherent drive with rate amplitude $\alpha$ from the left direction as an input to the system, the input-output relations are obtained as:
\begin{subequations}
\begin{align}
    \label{seq:input_output}
    a_\text{out}^R &= a_\text{in}^L + \sqrt{\gamma}S_- = \alpha + \sqrt{\gamma}S_- \\
    a_\text{out}^L &= a_\text{in}^R + \sqrt{\gamma}S_- = \sqrt{\gamma}S_-
\end{align}
\end{subequations}
where the operators $a_\text{out}^{R/L}$ define the transmitted/reflected field corresponding to the coherent drive determined as $a_\text{in}^L\ket{\alpha} = \alpha\ket{\alpha}$. We assume no input state from the right, $a_\text{in}^R = 0$. When atoms are separated by half-wavelength distances, the collective atomic dipole operators are instead defined as a sum of individual dipole operators with alternating signs, $S_\pm = \sum_{i=1}^N (-1)^{i} S_{\pm,i}$, accounting for the phase difference between adjacent atoms. The study of atoms with higher energy levels and individual atomic coupling to the waveguide is out of scope for this work, for a more general case see \cite{Lalumire2013}.

In the steady state, the output photon flux in the transmitted and reflected fields is equal to the input photon flux $|\alpha|^2 = 
\bexpval{a_\text{in}^{L\dag} a_\text{in}^L} =\bexpval{a_\text{out}^{R\dag} a_\text{out}^R} + \bexpval{a_\text{out}^{L\dag} a_\text{out}^L}$, also known as the balance relation. The average contribution of the transmitted field is obtained as,
\begin{equation}
\begin{split}
\label{seq:avg_transmitted_field}
    \bexpval{a_\text{out}^{R\dag} a_\text{out}^R} &= \bexpval{\big(a_\text{in}^{L\dag} + \sqrt{\gamma}S_+\big)\big(a_\text{in}^L + \sqrt{\gamma}S_-\big)} = \bexpval{a_\text{in}^{L\dag} a_\text{in}^L} + \sqrt{\gamma}\bexpval{a_\text{in}^{L\dag}S_-} + \sqrt{\gamma}\bexpval{a_\text{in}^{L}S_+} + \gamma \bexpval{S_+S_-\vphantom{a^L}} \\
    &= |\alpha|^2 + \sqrt{\gamma}\alpha^*\bexpval{S_-} + \sqrt{\gamma}\alpha\bexpval{S_+} + \gamma \bexpval{S_+S_-\vphantom{a^L}}.
\end{split}
\end{equation}
and the contribution from the reflected field is $\bexpval{a_\text{out}^{L\dag} a_\text{out}^L} = \gamma \bexpval{S_+S_-}$. According to the balance relation,
\begin{equation}
    |\alpha|^2 = |\alpha|^2 + \sqrt{\gamma}\alpha^*\bexpval{S_-} + \sqrt{\gamma}\alpha\bexpval{S_+} + 2\gamma \bexpval{S_+S_-\vphantom{a^L}},
\end{equation}
and by using \cref{seq:avg_S+,seq:avg_S+S-}, we obtain
\begin{equation}
    \sqrt{\gamma}(\alpha^*g^* + \alpha g) = -2\gamma |g|^2.
\end{equation}
Substituting $g=i \Omega / \gamma$ and solving for $\alpha$ gives us
\begin{equation}
    \Im(\alpha) = \frac{\Omega}{\sqrt{\gamma}}.
\end{equation}
We set $\alpha = i\Omega/\sqrt{\gamma}$ and the terms $\bexpval {a_{\text{in(out)}}^{R/L\dagger} a_{\text{in(out)}}^{R/L}}$ thus have the dimension of 1/time, describing the rate of photons in the waveguide.

\section{Scattered field}
\label{ssec:scattered_field}

The scattered photon flux of the atoms can be calculated using  \cref{seq:avg_S+,seq:avg_S+S-,seq:avg_transmitted_field}. The average reflected and transmitted photon flux is
\begin{subequations}
\begin{align}
    \bar{n}_\text{ref} &= \bexpval{a_\text{out}^{L\dag} a_\text{out}^L} = \gamma |g|^2 \left(1-\frac{N+1}{D}\right), \\
    \bar{n}_\text{trans} &= \bexpval{a_\text{out}^{R\dag} a_\text{out}^R} = \gamma |g|^2 \frac{N+1}{D}.
\end{align}
\end{subequations}
The scattered photon flux consists partly of the phase coherent components,
\begin{subequations}
\begin{align}
    \bar{n}_\text{ref,coh} &= \left|\bexpval{a_\text{out}^{L}}\right|^2 = \gamma |g|^2 \left(1-\frac{N+1}{D}\right)^2 , \\
    \bar{n}_\text{trans,coh} &= \left|\bexpval{a_\text{out}^{R}}\right|^2 = \gamma |g|^2 \left(\frac{N+1}{D}\right)^2,
\end{align}
\end{subequations}
and the phase incoherent components, $\bar{n}_\text{ref/trans,inc} = \bar{n}_\text{ref/trans}-\bar{n}_\text{ref/trans,coh}$,
\begin{equation}
\begin{split}
    \bar{n}_\text{ref,inc} = \bar{n}_\text{trans,inc} = \gamma |g|^2 \frac{N+1}{D}\left(1-\frac{N+1}{D}\right).
\end{split}
\end{equation}
In the regime of weak driving, $|g| \ll 1$, using $D \sim (N!)^2|g|^{-2N}$ from \cref{seq:D_weak_drive}, we obtain that most of the input photons are coherently reflected while transmission is rare and predominantly incoherent 
\begin{subequations}
\label{seq:weak_power}
\begin{align}
    \bar{n}_\text{ref} &= \gamma |g|^2 + \mathcal{O}\big(|g|^{2(N+1)}\big), \\
    \bar{n}_\text{trans} &= \gamma |g|^{2(N+1)} \frac{N+1}{(N!)^2}  + \mathcal{O}\big(|g|^{2(N+2)}\big), \label{seq:weak_power_trans} \\
    \bar{n}_\text{ref,coh} &= \gamma |g|^2 + \mathcal{O}\big(|g|^{2(N+1)}\big), \\
    \bar{n}_\text{trans,coh} &= \gamma |g|^{2(2N+1)} \frac{(N+1)^2}{(N!)^4} + \mathcal{O}\big(|g|^{2(2N+2)}\big), \\
    \bar{n}_\text{ref/trans,inc} &= \gamma |g|^{2(N+1)} \frac{N+1}{(N!)^2} + \mathcal{O}\big(|g|^{2(N+2)}\big), \label{seq:weak_power_trans_inc}
\end{align}
\end{subequations}
The expressions in \cref{seq:weak_power_trans,seq:weak_power_trans_inc} give the result in Eq. (3) of the main text. In the regime of strong driving, $|g| \gg 1$, keeping terms up to $\mathcal{O}(|g|^{-2})$ in the normalization constant $D \sim (N+1) + |g|^{-2} (N+2)(N+1)N/6$, we see that most of the input photons are coherently transmitted
\begin{subequations}
\label{seq:strong_power}
\begin{align}
    \bar{n}_\text{ref}  &= \gamma \frac{N(N+2)}{6} + \mathcal{O}\big(|g|^{-2}\big), \\
    \bar{n}_\text{trans} &= \gamma |g|^2 + \mathcal{O}\big(1\big), \\
    \bar{n}_\text{ref,coh} &= \gamma \bigg(\frac{(N+2)N}{6|g|}\bigg)^2 + \mathcal{O}\big(|g|^{-4}\big), \\
    \bar{n}_\text{trans,coh} &= \gamma |g|^2 + \mathcal{O}\big(1\big),  \\
    \bar{n}_\text{ref/trans,inc} &= \gamma \frac{N(N+2)}{6} + \mathcal{O}\big(|g|^{-2}\big).
\end{align}
\end{subequations}

\section{Scattered field: Higher order correlation functions}
\label{ssec:scattered_field_higher}

In this section, the steady-state solution in \cref{seq:steady_state} is used to calculate all higher order correlation functions for the reflected and transmitted field in both the weak and strong driving regime.

\subsection{Reflected field}
The correlation functions for the reflected field were first calculated by Drummond in \cite{Drummond1980}, but will also be included here for completeness. The $n$-th order correlation function for the reflected field is given by
\begin{equation}
\label{seq:Gn_r}
    G^{(n)}_\text{ref}(0) = \bexpval{\big(a_\text{out}^{L\dag}\big)^n \big(a_\text{out}^L\big)^n} = \gamma^n \bexpval{S_+^n S_-^n}.
\end{equation}
For $n>N$, it is obvious that $G^{(n)}_\text{ref}(0) = 0$ since the atomic system can only reflect up to $N$ photons simultaneously. Now we proceed to the case of $n \leq N$. 
\subsubsection{Strong driving regime}
In the strong driving regime, we have that $D \sim N+1$ from \cref{seq:D_strong_drive}, and the leading-order term in \cref{seq:Gn_r} is given by setting $r=n$ in \cref{seq:avg_S+nS-n}, and since $H_{n} = \binom{N+n+1}{2n+1}(n!)^2$, we obtain 
\begin{equation}
    G^{(n)}_\text{ref}(0) = \gamma^n \binom{N+n+1}{2n+1} \frac{(n!)^2}{N+1} + \mathcal{O}\big(|g|^{-2}\big), \quad |g| \gg 1.
\end{equation}
This gives the coherence function of the reflected field
\begin{equation}
    g^{(n)}_\text{ref}(0) = \frac{G^{(n)}_\text{ref}(0)}{[G^{(1)}_\text{ref}(0)]^n} = \binom{N+n+1}{2n+1} \frac{(n!)^2}{N+1} \left[\frac{6}{N(N+2)}\right]^n, \quad |g| \gg 1.
\end{equation}
\subsubsection{Weak driving regime}
For weak external field driving, we have that $D \sim (N!)^2 |g|^{-2N}$ from \cref{seq:D_weak_drive}, and the leading-order term is given by setting $r=N$ in \cref{seq:avg_S+nS-n}, and since $H_{N} = (N!)^2$, we obtain 
\begin{equation}
    G^{(n)}_\text{ref}(0) = \gamma^n |g|^{2n} + \mathcal{O}\big(|g|^{2(N+1)}\big), \quad |g| \ll 1,
\end{equation}
and the coherence function of the reflected field
\begin{equation}
    g^{(n)}_\text{ref}(0) = 1, \quad |g| \ll 1.
\end{equation}

\subsection{Transmitted field}
The $n$-th order correlation function for the transmitted field is given by
\begin{equation}
    G^{(n)}_\text{trans}(0) = \bexpval{\big(a_\text{out}^{R\dag}\big)^n \big(a_\text{out}^R\big)^n} = \bexpval{\big(\alpha^* + \sqrt{\gamma}S_+\big)^n \big(\alpha + \sqrt{\gamma}S_-\big)^n} = \gamma^n \bexpval{\big(g^* + S_+\big)^n \big(g + S_-\big)^n},
\end{equation}
where $\alpha = \sqrt{\gamma}g$ is used in the last step. Using the binomial theorem, $(x+y)^n = \sum _{k=0}^{n}\binom{n}{k}x^{n-k}y^{k}$, we expand to
\begin{equation}
\begin{split}
\label{seq:expanded_correlation_function}
    G^{(n)}_\text{trans}(0) &= \gamma^n \expval{\sum_{k=0}^n \binom{n}{k} (g*)^{n-k} S_+^k \sum_{j=0}^n \binom{n}{j} g^{n-j} (-S_-)^j } \\
    &= \gamma^n |g|^{2n} \sum_{k,j=0}^n \binom{n}{k} \binom{n}{j} (g^*)^{-k} g^{-j} \bexpval{S_+^kS_-^j}.
\end{split}
\end{equation}
Substituting the expectation value with the expression from \cref{seq:avg_S+nS-m} gives that,
\begin{equation}
\begin{split}
\label{seq:further_expanded_correlation_function}
    G^{(n)}_\text{trans}(0) &= \frac{\gamma^n |g|^{2n}}{D} \sum_{k,j=0}^n \binom{n}{k} \binom{n}{j}(-1)^{k+j} \sum_{r = \max(k,j)}^N |g|^{-2r} H_{r},
\end{split}
\end{equation}
with $n \leq N$ and where we have used that $g^* = -g$.

\subsubsection{Strong driving regime}
Once again, for strong external field driving, we have that $D \sim N+1$ from \cref{seq:D_strong_drive}, and the leading-order term is given by setting $k=j=r=0$ in \cref{seq:further_expanded_correlation_function}, and since $H_{0} = \frac{(N+1)!}{N!} = N+1$, we obtain
\begin{equation}
    G^{(n)}_\text{trans}(0) = \gamma^n |g|^{2n} + \mathcal{O}\big(|g|^{2(n-1)}\big), \quad |g| \gg 1.
\end{equation}
and the coherence function of the transmitted field
\begin{equation}
    g^{(n)}_\text{trans}(0) = 1, \quad |g| \gg 1.
\end{equation}

\subsubsection{Weak driving regime}
\label{ssec:scattered_field_higher_trans_weak}

The leading-order term for weak driving is obtained for the largest $r$ in \cref{seq:further_expanded_correlation_function}. At first glance, it seems that $r=N$ would give us the leading-order term but then the outer summations sum to zero. To proceed, we instead notice that
\begin{equation}
    \sum_{r=\max(k,j)}^N |g|^{-2r} H_{r} = \sum_{r=0}^N |g|^{-2r} H_{r} -  \sum_{r=0}^{\max(k,j)-1} |g|^{-2r} H_{r} = D -  \sum_{r=0}^{\max(k,j)-1} |g|^{-2r} H_{r}.
\end{equation}
and substituting this into \cref{seq:further_expanded_correlation_function} yields
\begin{equation}
\label{seq:even_further_expanded_correlation_function}
    G^{(n)}_\text{trans}(0) = \frac{\gamma^n |g|^{2n}}{D} \sum_{k,j=0}^n \binom{n}{k} \binom{n}{j}(-1)^{k+j} \left(D -  \sum_{r=0}^{\max(k,j)-1} |g|^{-2r} H_{r}\right).
\end{equation}
The first part of the summands in \cref{seq:even_further_expanded_correlation_function}, which do not depend on $r$, are now shown to sum to zero
\begin{equation}
\begin{split}
    \sum_{k,j=0}^n \binom{n}{k} \binom{n}{j} (-1)^{k+j} = \left(\sum_{k=0}^n \binom{n}{k} (-1)^{k}\right)^2 = (1-1)^{2n}  = 0,
\end{split}
\end{equation}
where the binomial theorem is used in the second to last step. This leaves us with
\begin{equation}
    G^{(n)}_\text{trans}(0) = -\frac{\gamma^n |g|^{2n}}{D} \sum_{k,j=0}^n \binom{n}{k} \binom{n}{j}(-1)^{k+j}\sum_{r=0}^{\max(k,j)-1} |g|^{-2r} H_{r}.
\end{equation}
For $n \leq N$, we see that the leading-order term for weak driving is obtained when $r=\max(k,j)-1=n-1$, which happens either when: (i) $k=j=n$, (ii) or $k=n$, and $j<n$, (iii) or $j=n$ and $k<n$. These three cases, and substituting with $D \sim  (N!)^2 |g|^{-2N}$ from \cref{seq:D_weak_drive} for weak driving, give us
\begin{equation}
\label{seq:G_n_trans_weak}
\begin{split}
    G^{(n)}_\text{trans}(0) &= -\frac{\gamma^n |g|^{2(N+1)}}{(N!)^2} H_{n-1} \left( \binom{n}{n}^2 (-1)^{2n}  + 2(-1)^{n} \sum_{k=0}^{n-1} \binom{n}{k} (-1)^{k} \right) + \mathcal{O}\big(|g|^{2(N+2)}\big) \\
    &= -\frac{\gamma^n |g|^{2(N+1)}}{(N!)^2} \binom{N+n}{2n-1} (n-1)!^2 \left(1 + 2(-1)^n (-1)^{n-1}\right) + \mathcal{O}\big(|g|^{2(N+2)}\big) \\
    &= \gamma^n |g|^{2(N+1)} \binom{N+n}{2n-1} \left[\frac{(n-1)!}{N!}\right]^2 + \mathcal{O}\big(|g|^{2(N+2)}\big), \quad |g| \ll 1.
\end{split}
\end{equation}
The coherence function of the transmitted field is then
\begin{equation}
    \label{seq:g_n_trans}
    g^{(n)}_\text{trans}(0) = |g|^{-2(N+1)(n-1)} \binom{N+n}{2n-1} \frac{(n-1)!^2}{(N+1)^n} (N!)^{2(n-1)}, \quad |g| \ll 1.
\end{equation}
The second-order coherence function $g^{(2)}_\text{trans}(0)$ which is presented in the main text to show superbunched transmission is obtained by setting $n=2$ in \cref{seq:g_n_trans}. For the case where the order of the correlation function $n$ exceeds the number of atoms $N$, i.e., $n = N + m$ with $m \geq 1$, the upper bound of the first summations in \cref{seq:further_expanded_correlation_function} are modified such that they only run to up $N$,
\begin{equation}
\label{seq:further_expanded_correlation_function_N+m}
    G^{(N+m)}_\text{trans}(0) = \frac{\gamma^{N+m} |g|^{2(N+m)}}{D} \sum_{k,j=0}^N \binom{N+m}{k} \binom{N+m}{j}(-1)^{k+j} \sum_{r = \max(k,j)}^N |g|^{-2r} H_{r},
\end{equation}
since $\bexpval{S_+^k S_-^j} = 0$ for $k,j > N$. The leading-order term is then simply given for $r=N$ in \cref{seq:further_expanded_correlation_function_N+m}, and substituting with $D \sim  (N!)^2 |g|^{-2N}$ from \cref{seq:D_weak_drive} for weak driving gives us
\begin{equation}
\label{seq:G_N+m_trans_weak}
\begin{split}
    G^{(N+m)}_\text{trans}(0) &= \frac{\gamma^{N+m} |g|^{2(N+m)}}{(N!)^2} H_{N} \sum_{k,j=0}^N \binom{N+m}{k} \binom{N+m}{j}(-1)^{k+j}  + \mathcal{O}\big(|g|^{2(N+m+1)}\big) \\
    &= \gamma^{N+m} |g|^{2(N+m)} \left(\sum_{k=0}^N \binom{N+m}{k} (-1)^{k}\right)^2 + \mathcal{O}\big(|g|^{2(N+m+1)}\big) \\
    &= \gamma^{N+m} |g|^{2(N+m)} \binom{N+m-1}{N}^2 + \mathcal{O}\big(|g|^{2(N+m+1)}\big), \quad |g| \ll 1.
\end{split}
\end{equation}
The summation in the last step can be found in \cite{Riordan1968}. The results in \cref{seq:G_n_trans_weak,seq:G_N+m_trans_weak} give the correlation function in Eq. (4) of the main text. The coherence function of the transmitted field is then
\begin{equation}
    g^{(N+m)}_\text{trans}(0) = |g|^{-2N(N+m)} \binom{N+m-1}{N}^2 \frac{(N!)^{2(N+m)}}{(N+1)^{N+m}}, \quad |g| \ll 1.
\end{equation}

\section{No-click Lindblad master equation}
The Lindblad master equation \cite{Molmer1993, Breuer2007, Wiseman2009} for a system with density matrix $\rho$, Hamiltonian $H$ and a set of jump operators $\mathbf{L}$ is ($\hbar=1$):
\begin{equation}
    \dot{\rho} = -i[H,\rho] + \sum_{L\in\mathbf{L}} \mathcal{D}(L)\rho,
\end{equation}
where $\mathcal{D}(L)\rho = \mathcal{D}(L) = L\rho L^\dag -\tfrac{1}{2}\{L^\dag L, \rho\}$ is the Lindblad dissipator. This dissipator consists of two terms that can be interpreted as a click/jump and drift term,
\begin{equation}
    \mathcal{D}(L) = \underbrace{\vphantom{\tfrac{1}{2}}L\rho L^\dag}_{\text{click}} \underbrace{-\tfrac{1}{2}\{L^\dag L, \rho\}}_{\text{drift}}.
\end{equation}
In our case, the click term describes a photon detection while the drift term describes the system dynamics between detection events. The no-click master equation then consists of at least one Lindblad dissipator without the click term. To ensure trace-preserving dynamics, we define the remaining drift part of the dissipator as
\begin{equation}
    \mathcal{D}_\text{drift}(L) = -\tfrac{1}{2}\{L^\dag L, \rho\} + \Tr(L^\dag L \rho) \rho.
\end{equation}

\subsection{No-click in the transmitted field}
\label{ssec:no_click_trans}
The master equation for no clicks in the transmitted field is
\begin{equation}
    \dot{\rho} = -i \tfrac{\Omega}{2}[S_+ + S_-, \rho] + \mathcal{D}(a_\text{out}^L)+ \mathcal{D}_\text{drift}(a_\text{out}^R) = \mathcal{D}(a_\text{out}^R) + \mathcal{D}_\text{drift}(a_\text{out}^R),
\end{equation}
and expressing the drift part of the dissipator gives,
\begin{equation}
    \label{seq:master_no_transmitted_clicks}
    \dot{\rho} = 2\mathcal{D}(a_\text{out}^R) - a_\text{out}^R \rho a_\text{out}^{R\dag} + \Tr\big(a_\text{out}^{R\dag} a_\text{out}^R\rho\big)\rho=a_\text{out}^R \rho a_\text{out}^{R\dag}-a_\text{out}^{R\dag}a_\text{out}^R \rho-\rho a_\text{out}^{R\dag}a_\text{out}^R + \bexpval{a_\text{out}^{R\dag} a_\text{out}^R}\rho.
\end{equation}
Our analysis then focuses on the state of the system following the detection of a transmitted photon, which, while rare in the weak driving regime, is inevitable given sufficient time. The unnormalized state conditioned on detecting a transmitted photon is given by:
\begin{equation}
    \label{seq:unnormalized_conditional_state}
    \Tilde{\rho}_c = a_\text{out}^R \rho a_\text{out}^{R\dag} \quad \Longrightarrow \quad \rho =  (a_\text{out}^R)^{-1} \Tilde{\rho}_c (a_\text{out}^{R\dag})^{-1}.
\end{equation} 
Substituting for this conditional state in \cref{seq:master_no_transmitted_clicks} gives the conditional master equation
\begin{equation}
    \dot{\Tilde{\rho}}_c = a_\text{out}^R \Tilde{\rho}_c a_\text{out}^{R\dag}-a_\text{out}^R a_\text{out}^{R\dag} \Tilde{\rho}_c-\Tilde{\rho}_c a_\text{out}^R a_\text{out}^{R\dag} + \bexpval{a_\text{out}^{R\dag} a_\text{out}^R} \Tilde{\rho}_c.
\end{equation}
In the weak driving regime, we have that
\begin{equation}
    \frac{a_\text{out}^R}{\sqrt{\gamma}} = S_- + \mathcal{O}(|g|), \quad |g| \ll 1,
\end{equation}
and $\bexpval{a_\text{out}^{R\dag} a_\text{out}^R} \simeq 0$, giving the conditional master equation,
\begin{equation}
    \label{seq:no_ref_click_conditional_master}
    \frac{\dot{\Tilde{\rho}}_\text{c}}{\gamma} = S_- \Tilde{\rho}_\text{c} S_+-S_-S_+ \Tilde{\rho}_\text{c}-\Tilde{\rho}_\text{c} S_-S_+ + \mathcal{O}(|g|), \quad |g| \ll 1.
\end{equation}
To find the steady-state solution, i.e. $\dot{\Tilde{\rho}}_\text{c} = 0$, we substitute with the diagonal \emph{ansatz}
\begin{equation}
\label{seq:ansatz}
    \Tilde{\rho}_\text{c} = \sum_{k=0}^{N} p_k \ketbra{k},
\end{equation}
where $k$ denotes the number of atoms in the ground state, in \cref{seq:no_ref_click_conditional_master}, giving us
\begin{equation}
\label{seq:solvable_no_click_conditional_master}
\begin{split}
    0 &= \sum_{k=0}^N p_k \bigg[ S_- \ketbra{k} S_+ - S_- S_+ \ketbra{k} - \ketbra{k} S_- S_+\bigg] \\
    &= \sum_{k=0}^N p_k \bigg[(k+1)(N-k) \ketbra{k+1} -2k(N-k+1)\ketbra{k}\bigg]
\end{split}
\end{equation}
where the relations in \cref{seq:dipole_operators} were used in the last step. \cref{seq:solvable_no_click_conditional_master} now reveals the recurrence relation
\begin{equation}
    p_k (k+1)(N-k) -2p_{k+1}(k+1)(N-(k+1)+1) = 0 \quad \Longrightarrow \quad p_{k+1} = \frac{p_k}{2},
\end{equation}
for the probabilities $p_k$, with the solution
\begin{equation}
    p_k = \frac{p_0}{2^k}.
\end{equation}
Finally, the normalization requirement of the density matrix gives that
\begin{equation}
    \sum_{k=0}^N \frac{p_0}{2^k} = 1 \quad \Longrightarrow \quad p_0^{-1} = \sum_{k=0}^N \frac{1}{2^k} = 2-2^{-N}.
\end{equation}
The normalized steady-state solution of the conditional master equation from \cref{seq:no_ref_click_conditional_master} is then
\begin{equation}
\label{seq:ss_no_ref_click_conditional_master}
    \rho_\text{c} = \sum_{k=0}^N \frac{1}{2^k(2-2^{-N})} \ketbra{k} = \sum_{k'=0}^N \frac{1}{2^{N-k'}(2-2^{-N})} \ketbra{k'}_\text{ex},
\end{equation}
where $\ket{k'}_\text{ex}$ is the state with $k'$ excited atoms. It is worth noting that the probabilities of the conditional steady state in \cref{seq:ss_no_ref_click_conditional_master} follows a truncated geometric distribution with failure probability $1/2$. The conditional state result in \cref{seq:ss_no_ref_click_conditional_master} is presented in Eq. (6) of the main text.

The steady-state solution of the no-click master equation is then
\begin{equation}
\label{seq:ss_no_ref_click}
    \rho \propto (a_\text{out}^R)^{-1} \rho_\text{c} (a_\text{out}^{R\dag})^{-1} \propto \sum_{m=0}^N\left(\frac{S_-}{g^*}\right)^m \sum_{k=0}^{N} p_k \ketbra{k} \sum_{n=0}^N\left(\frac{S_+}{g}\right)^n.
\end{equation}
By using \cref{seq:dipole_operators} we have that
\begin{equation}
\begin{split}
    S_-^n \ket{p} = \sqrt{\frac{(p+n)!(N-p)!}{p!(N-p-n)!}}\ket{p+n}, \quad 0\leq n \leq N-p,
\end{split}
\end{equation}
and applying this to \cref{seq:ss_no_ref_click} gives the steady-state solution
\begin{equation}
\begin{split}
    \rho &\propto \sum_{\substack{0\leq k \leq N \\ 0\leq m,n \leq N-k}} p_k (g^*)^{-m} (g)^{-n} \sqrt{\frac{(k+m)!(k+n)!}{(N-k-m)!(N-k-n)!}} \frac{(N-k)!}{k!} \ketbra{k+m}{k+n} \\ &\approx \ketbra{N} = \ketbra{0}_\text{ex}, \quad |g| \ll 1.
\end{split}
\end{equation}
We see that, in the weak driving regime, the probability of the system being in the fully excited state is independent of $|g|$. In contrast, the probability of the system being in the ground state is to leading-order proportional to $|g|^{-2N}$, which is exponentially larger than the probability of being in the fully excited state for $|g| \ll 1$. The average excitation number in the system after detecting a transmitted photon is then,
\begin{equation}
\label{seq:avg_excitations}
    \sum_{k=0}^N \frac{(N-k)}{2^k(2-2^{-N})} = \frac{2^{N+1}N+1}{2^{N+1}-1} - 1,
\end{equation}
which for large $N$ (already at 4) converges to $N-1$.

\subsection{No-click in the reflected and transmitted field}
The master equation for no clicks in both the reflected and transmitted field is
\begin{equation}
    \label{seq:no_click_master}
    \dot{\rho} = -i \tfrac{\Omega}{2}[S_+ + S_-, \rho] + \mathcal{D}_\text{drift}(a_\text{out}^L) + \mathcal{D}_\text{drift}(a_\text{out}^R).
\end{equation}
By using that,
\begin{equation}
    \mathcal{D}(a_\text{out}^R) = -i \tfrac{\Omega}{2}[S_+ + S_-, \rho] + \mathcal{D}(a_\text{out}^L) = -i \tfrac{\Omega}{2}[S_+ + S_-, \rho] + \mathcal{D}_\text{drift}(a_\text{out}^L) + \mathcal{D}_\text{jump}(a_\text{out}^L),
\end{equation}
the master equation in \cref{seq:no_click_master} is rewritten as
\begin{equation}
\label{seq:further_no_click_master}
\begin{split}
    \dot{\rho} &= \mathcal{D}(a_\text{out}^R) -  \mathcal{D}_\text{jump}(a_\text{out}^L) + \mathcal{D}_\text{drift}(a_\text{out}^R) \\
    &= 2\mathcal{D}_\text{drift}(a_\text{out}^R) + \mathcal{D}_\text{jump}(a_\text{out}^R) - \mathcal{D}_\text{jump}(a_\text{out}^L) \\
    & = a_\text{out}^{R} \rho a_\text{out}^{R\dag}  - a_\text{out}^{R\dag} a_\text{out}^R \rho - \rho a_\text{out}^{R\dag} a_\text{out}^R - a_\text{out}^{L} \rho a_\text{out}^{L\dag} + \Tr\big(a_\text{out}^{L\dag} a_\text{out}^L \rho\big) \rho + \Tr\big(a_\text{out}^{R\dag} a_\text{out}^R \rho\big)\rho \\
    & = a_\text{out}^{R} \rho a_\text{out}^{R\dag}  - a_\text{out}^{R\dag} a_\text{out}^R \rho - \rho a_\text{out}^{R\dag} a_\text{out}^R - a_\text{out}^{L} \rho a_\text{out}^{L\dag} + |\alpha|^2 \rho.
\end{split}
\end{equation}
The relation $\bexpval{a_\text{out}^{R\dag} a_\text{out}^R} + \bexpval{a_\text{out}^{L\dag} a_\text{out}^L} = \bexpval{a_\text{in}^{L\dag} a_\text{in}^L} = |\alpha|^2$ was used in the last step. Substituting with the unnormalized state conditioned on detected transmitted photon from \cref{seq:unnormalized_conditional_state} in \cref{seq:further_no_click_master} gives us
\begin{equation}
\begin{split}
    (a_\text{out}^R)^{-1} \dot{\Tilde{\rho}}_c (a_\text{out}^{R\dag})^{-1} &= \Tilde{\rho}_c - a_\text{out}^{R\dag} \Tilde{\rho}_c (a_\text{out}^{R\dag})^{-1} - (a_\text{out}^R)^{-1} \Tilde{\rho}_c a_\text{out}^R - a_\text{out}^{L} (a_\text{out}^R)^{-1} \Tilde{\rho}_c (a_\text{out}^{R\dag})^{-1} a_\text{out}^{L\dag} \\ 
    & \qquad + |\alpha|^2 (a_\text{out}^R)^{-1} \Tilde{\rho}_c (a_\text{out}^{R\dag})^{-1}.
\end{split}
\end{equation}
Multiplying with the transmitted output operator and its conjugate on both sides and utilizing the fact that the output operators commute with each other gives the conditional master equation
\begin{equation}
\begin{split}
    \dot{\Tilde{\rho}}_c &= a_\text{out}^{R} \Tilde{\rho}_c a_\text{out}^{R\dag} - a_\text{out}^{R}a_\text{out}^{R\dag} \Tilde{\rho}_c -  \Tilde{\rho}_c a_\text{out}^R a_\text{out}^{R\dag} - a_\text{out}^{L}  \Tilde{\rho}_c  a_\text{out}^{L\dag} + |\alpha|^2 \Tilde{\rho}_c \\
    &= |\alpha|^2 \Tilde{\rho}_c + \alpha^*\sqrt{\gamma}S_- \Tilde{\rho}_c + \alpha \sqrt{\gamma} \Tilde{\rho}_c S_+ + \gamma S_- \Tilde{\rho}_c S_+
    - |\alpha|^2 \Tilde{\rho}_c - \alpha^*\sqrt{\gamma}S_-  \Tilde{\rho}_c  - \alpha \sqrt{\gamma} S_+  \Tilde{\rho}_c - \gamma S_- S_+  \Tilde{\rho}_c \\
    &\quad- |\alpha|^2 \Tilde{\rho}_c - \alpha^*\sqrt{\gamma} \Tilde{\rho}_c S_- - \alpha \sqrt{\gamma} \Tilde{\rho}_c S_+ - \gamma \Tilde{\rho}_c S_- S_+
    - \gamma S_- \Tilde{\rho}_c S_+ + |\alpha|^2 \Tilde{\rho}_c \\
    &= - \alpha \sqrt{\gamma} S_+  \Tilde{\rho}_c - \gamma S_- S_+  \Tilde{\rho}_c - \alpha^*\sqrt{\gamma} \Tilde{\rho}_c S_- - \gamma \Tilde{\rho}_c S_- S_+ \\
    &= -i \Omega (S_+  \Tilde{\rho}_c  - \Tilde{\rho}_c S_-) - \gamma(S_- S_+  \Tilde{\rho}_c + \Tilde{\rho}_c S_- S_+).
\end{split}
\end{equation}
The conditional steady-state solution is then straightforwardly identified as
\begin{equation}
\label{seq:ss_no_click_conditional_master}
    \Tilde{\rho}_c = S_+^N S_-^N,
\end{equation}
while recalling that $S_+^{N+1} = S_-^{N+1} = 0$. Finally, the trace of this state is calculated
\begin{equation}
\begin{split}
    \Tr(S_+^N S_-^N) &= \sum_{k=0}^N \mel{k}{S_+^N S_-^N}{k} = \mel{0}{S_+^N S_-^N}{0} = (N!)^2
\end{split}
\end{equation}
which gives the normalized conditional steady-state solution,
\begin{equation}
\label{seq:ss_no_click_conditional_master_norm}
    \rho_c = \frac{S_+^NS_-^N}{(N!)^2} = \ketbra{0} = \ketbra{N}_\text{ex},
\end{equation}
i.e., the fully excited state. The result in \cref{seq:ss_no_click_conditional_master_norm} is presented in the conditional atomic state section of the main text.

The steady-state solution of \cref{seq:no_click_master} is then,
\begin{equation}
\label{seq:ss_no_click}
\begin{split}
    \rho &\propto (a_\text{out}^R)^{-1} \rho_\text{c} (a_\text{out}^{R\dag})^{-1} \\ &
    \propto \sum_{m=0}^N\left(\frac{S_-}{g^*}\right)^m \ketbra{0} \sum_{n=0}^N\left(\frac{S_+}{g}\right)^n \\
    &= \sum_{m,n=0}^N (g^*)^{-m} (g)^{-n} \sqrt{\frac{m!n!}{(N-m)!(N-n)!}} \ketbra{m}{n} \\
    &\approx \ketbra{N} = \ketbra{0}_\text{ex}, \quad |g| \ll 1.
\end{split}
\end{equation}
i.e., a state very close to the ground state in the weak driving regime. Interestingly, since the transmitted and reflected output operators, $a_\text{out}^R$ and $a_\text{out}^L$, commute, we see that if we instead condition the steady state on $k$ reflected clicks followed by a transmitted click, we obtain the state
\begin{equation}
\begin{split}
    \rho_c &\propto a_\text{out}^R (a_\text{out}^L)^k \rho (a_\text{out}^{L\dag})^k a_\text{out}^{R\dag} \\
    &= a_\text{out}^R (a_\text{out}^L)^k (a_\text{out}^R)^{-1} S_+^N S_-^N (a_\text{out}^{R\dag})^{-1} (a_\text{out}^{L\dag})^k a_\text{out}^{R\dag} \\
    &= (a_\text{out}^L)^k a_\text{out}^R (a_\text{out}^R)^{-1} S_+^N S_-^N (a_\text{out}^{R\dag})^{-1} a_\text{out}^{R\dag} (a_\text{out}^{L\dag})^k \\
    &= (a_\text{out}^L)^k S_+^N S_-^N (a_\text{out}^{L\dag})^k \\
    &\propto S_-^k S_+^N S_-^N S_+^k \\
    &\propto \ketbra{k} \\
    &= \ketbra{N-k}_\text{ex},
\end{split}
\end{equation}
i.e, the state with $N-k$ excitations. This means that our atomic system will be in a state with $N-k$ excitations. These excitations eventually decay into the waveguide, summing up to a total of $N+1$ photons.

\section{Photon bunching characteristics}
\subsection{Timescale of concurrent \texorpdfstring{$N+1$}{N+1} photon arrival enabling transmission}
\label{ssec:timescale_in}
This section estimates the timescale $T_\text{in}$ within which $N+1$ photons must arrive to enable transmission. For a coherent drive with rate amplitude $\alpha$, the number of incoming photons during the time $T_\text{in}$ is a Poisson random variable $X \sim \Pois(|\alpha|^2 T_\text{in})$ with an average of $|\alpha|^2 T_\text{in}$ photons. In the weak driving regime, we can safely omit photon bunching events consisting of more than $N+1$ photons. The probability of an incoming $N+1$-photon state during a time frame of length $T_\text{in}$ is then
\begin{equation}
    \Pr(X = N+1) = \frac{(|\alpha|^2 T_\text{in})^{N+1}}{(N+1)!}e^{-|\alpha|^2 T_\text{in}} \approx \frac{(|\alpha|^2 T_\text{in})^{N+1}}{(N+1)!},
\end{equation}
where weak driving is assumed in the last step. During this time frame, the probability per second of an incoming $(N+1)$-photon state is $\Pr(X = N+1)/T_\text{in}$. From \cref{seq:avg_excitations}, we have that after the first transmitted photon, the atomic system on average consists of $N-1$ excitations. As these excited atoms decay, they re-emit photons into the waveguide. On average, $(N+1)/2$ photons are transmitted, including the initially detected transmitted photon. In the weak driving regime, the average transmission rate of a $(N+1)$-photon bunch is then
\begin{equation}
\label{seq:transmission_rate_bunch}
     \frac{\Pr(X = N+1)}{T_\text{in}} \frac{N+1}{2} = \frac{|\alpha|^{2(N+1)}T_\text{in}^{N}}{2N!}.
\end{equation}
In the weak driving regime, we have that transmission is dominated by $(N+1)$-photon bunches, and we can set the rate in \cref{seq:transmission_rate_bunch} equal to the transmission rate in the weak driving regime from \cref{seq:weak_power_trans},
\begin{equation}
    \frac{|\alpha|^{2(N+1)}T_\text{in}^{N}}{2N!} = \gamma |g|^{2(N+1)} \frac{N+1}{(N!)^2}.
\end{equation}
By recalling that $|\alpha| = |g|\sqrt{\gamma} = \Omega/\sqrt{\gamma}$, we obtain
\begin{equation}
\begin{split}
\label{seq:T_in}
    T_\text{in} = \frac{1}{\gamma}\left(\frac{2(N+1)}{N!}\right)^{1/N} .
\end{split}
\end{equation}
For moderately large $N$, Stirling's formula, $N! \sim \sqrt {2\pi N}\left(\frac{N}{e}\right)^{N}$ gives the asymptote $T_\text{in} \sim \frac{e}{\gamma N}$. The timescale $T_\text{in}$ scales inversely with the number of atoms $N$, demonstrating that the photon bunching dynamics becomes faster as the number of atoms in the waveguide increases, a signature of collective behavior. The timescale in \cref{seq:T_in} is the result presented in Eq. (7) of the main text.

\subsection{Timescale of relaxation}
We define the timescale $T_\text{out}$ which characterizes how quickly a fully excited atomic state decays into the waveguide. Assuming no external driving, from the fully excited state $\ket{N}_\text{ex}$, the atomic system undergoes cascaded emission, transitioning sequentially from $\ket{k}_\text{ex}$ to $\ket{k-1}_\text{ex}$ at rates 
\begin{equation}
    \Gamma_{k} = 2\gamma \mel{k}{S_+ S_-}{k}_\text{ex} = 2\gamma(N-k+1)k,
\end{equation}
until reaching the ground state $\ket{0}_\text{ex}$. The average relaxation time \cite{Gross1982} is then,
\begin{equation}
    T_\text{out,avg} = \sum_{k=1}^{N} \frac{1}{\Gamma_k} = \frac{1}{2\gamma}\sum_{k=1}^{N} \frac{1}{(N-k+1)k} = \frac{1}{\gamma(N+1)} \sum_{k=1}^N \frac{1}{k} = \frac{H_N}{\gamma(N+1)} \sim \frac{\ln(N)}{\gamma N}
\end{equation}
where the last step is the asymptotic limit for large $N$ and $H_N$ is the $N$-th Harmonic number.

The cascaded emission process can also be described as a continuous-time absorbing Markov chain with $N+1$ states, where decay times between adjacent states follow exponential distributions with rates $\Gamma_k$. Thus, the relaxation time $T_\text{out}$ follows the hypoexponential distribution \cite{Amari1997, Bolch2006, Legros2015}, which is the sum of independent exponential random variables. The cumulative distribution function for the hypoexponential distribution \cite{Legros2015}, with rate parameters $\Gamma_N, \dots,  \Gamma_1$, is given by
\begin{equation}
    F(T_\text{out}) = 1-\boldsymbol{\alpha}\,e^{T_\text{out}\mathbf{Q}}\,\mathbf{1},
\end{equation}
where $T_\text{out}\geq0$, $\boldsymbol{\alpha}=[1,0,\dots,0]$ is a row vector, and $\mathbf{1}$ is a column vector of ones, both of length $N$. The transition-rate matrix $\mathbf{Q}$, of dimension $N \times N$, is defined as
\begin{equation}
    \mathbf{Q} = \begin{bmatrix}
        -\Gamma_N & \Gamma_N & 0 & \dots & 0 \\
        0 & -\Gamma_{N-1} & \Gamma_{N-1} & \ddots & \vdots \\
        \vdots & \ddots & \ddots & \ddots & 0 \\
        0 & 0 & \ddots & -\Gamma_2 & \Gamma_2 \\
        0 & 0 & \dots & 0 & -\Gamma_1
    \end{bmatrix}.
\end{equation}
The $p$-th relaxation time quantile $T_{\text{out},p}$ is the time at which a fully excited atomic system reaches the ground state with probability $p$, and is given by $T_{\text{out},p} = F^{-1}(p)$.

\subsection{Distribution of photon bunches}
\label{ssec:photon_bunches_dist}
This section aims to explore the probability distribution of photon bunches for a given input drive rate amplitude $\alpha = i\tfrac{\Omega}{\sqrt{\gamma}}$. An atomic system with $N$ atoms effectively acts as a mirror, coherently reflecting photons up to $N$ incoming photons, while transmission becomes possible for $N+1$ or more concurrent photons. The question is then, if we detect a photon in the transmitted field, what is the probability that it was exactly $N+1$ incoming photons from the drive and not more? If we detect a photon in the transmitted field, then the probability of the input photon state consisting of $k \in \mathbb{N}$, with $k \geq N+1$, photons follows an $N$-truncated Poisson distribution with the random variable $Y$ such that
\begin{equation}
    \Pr(Y=k) = \frac{\Pr(X=k)}{\Pr(X \geq N+1)},
\end{equation}
recalling the definition of $X$ from \cref{ssec:timescale_in}. To calculate the probability of a given sized photon bunch, we must consider the finite probability of additional photons being inputted from the drive during the time $T_\text{out}$ that it takes for our initial photons to be absorbed and emitted back into the waveguide. The probability of obtaining an $(N+1)$-photon bunch is then
\begin{equation}
    p(N+1) = \Pr(Y=N+1)\Pr(Z=0),
\end{equation}
where $Z \sim \Pois(|\alpha|^2 T_\text{out})$. This is the probability that the photon bunch is part of an input $N+1$ photon state multiplied by the probability that no further photons are inputted from the drive during the time it takes for the atoms to absorb and emit the photons. The choice of $T_\text{out}$ involves a trade-off: it must be long enough to allow the fully excited atomic system to relax, i.e., $T_\text{out} \gtrsim T_\text{out,99\%}$, but short enough to minimize excess photon input. Furthermore, the probability of an $(N+2)$-photon bunch must then be
\begin{equation}
 p(N+2) = \Pr(Y=N+2)\Pr(Z=0) + \Pr(Y=N+1)\Pr(Z=1).
\end{equation}
In general, we have that the probability of obtaining a photon bunch, conditioned on a transmitted click, of size $k$ is
\begin{equation}
\label{seq:photon_bunch_prob}
    p(k) = \sum_{i=0}^{k-N-1} \Pr(Y=k-i)\Pr(Z=i),
\end{equation}
where $k \in \mathbb{N}$ and $k \geq N+1$. It is easy to see that $p(k)$ defines a probability mass function since $p(k) \geq 0$ and
\begin{equation}
\begin{split}
    \sum_{k=N+1}^{\infty} p(k) &= \sum_{k=N+1}^{\infty} \sum_{i=0}^{k-N-1} \Pr(Y=k-i)\Pr(Z=i) \\
    &= \sum_{k=0}^\infty \sum_{i=N+1}^\infty \Pr(Y=k+N+1) \Pr(Z=i-(N+1)) \\
    &=\sum_{k=N+1}^\infty \Pr(Y=k) \sum_{i=0}^\infty \Pr(Z=i) \\
    &= 1.
\end{split}
\end{equation}
The result in \cref{seq:photon_bunch_prob} is presented in Eq. (8) of the main text.

The probability of an $(N+1)$-photon bunch can be bounded in the weak driving regime, by first rewriting
\begin{equation}
    p(N+1) = \frac{\Pr(X=N+1)}{\Pr(X \geq N+1)} \Pr(Z=0) = \frac{\Pr(X=N+1)\Pr(Z=0)}{\Pr(X = N+1) + \Pr(X \geq N+2)}  = \frac{\Pr(Z=0)}{1+\frac{\Pr(X \geq N+2)}{\Pr(X=N+1)}},
\end{equation}
and then using the bound,
\begin{equation}
    \Pr(X=N+2) = \frac{\lambda^{N+2} e^{-\lambda}}{(N+2)!} < \frac{(e\lambda)^{N+2}e^{-\lambda}}{\sqrt{2\pi(N+2)}(N+2)^{N+2}} < \Pr(X \geq N+2) \leq \frac{(e\lambda)^{N+2}e^{-\lambda}}{(N+2)^{N+2}},
\end{equation}
where Stirling's formula is used for the lower bound and a Chernoff bound is used for the upper bound \cite{Mitzenmacher2005}, to obtain
\begin{equation}
    \frac{e^{-|\alpha|^2T_\text{out}}}{1+\frac{|\alpha|^2T_\text{in}\sqrt{2\pi(N+2)}}{N+2}} < p(N+1) < \frac{e^{-|\alpha|^2T_\text{out}}}{1+\frac{|\alpha|^2T_\text{in}}{N+2}},
\end{equation}
for $\lambda = |\alpha|^2T_\text{in} < N+2$. We can then see that $p(N+1) \approx e^{-|\alpha|^2T_\text{out}}$ which is approximately one when $|\alpha|^2 T_\text{out} \ll 1$.

\bibliographystyle{apsrev4-2}